\documentclass{elsart}

\def\COBE{{\sl COBE}}
\def\MAP{{\sl MAP}}
\def\Planck{{\sl Planck}}

\def\fraction#1/#2{\leavevmode\kern.1em
 \raise.5ex\hbox{\the\scriptfont0 #1}\kern-.1em
 /\kern-.15em\lower.25ex\hbox{\the\scriptfont0 #2}}

\usepackage{harvard,graphics,amssymb}

\usepackage{amssymb}

\def\url#1{{\ttfamily\def\/{/\discretionary{}{}{}}#1}}
\def\bibcode#1{(\texttt{#1})}

\begin{document}

\begin{frontmatter}
\journal{New Astronomy}
\title{Morphing the CMB:\\
a technique for interpolating power spectra}
\author{Kris Sigurdson\thanksref{emailks}},
\thanks[emailks]{ksigurds@sfu.ca}
\address{Department of Physics, Simon Fraser University,
    Burnaby, BC V5A 1S6, Canada}
\author{Douglas Scott\thanksref{emailds}\thanksref{corresp}}
\thanks[emailds]{dscott@astro.ubc.ca}
\thanks[corresp]{Corresponding author}
\address{Department of Physics and Astronomy, University of British
Columbia, Vancouver, BC V6T 1Z1, Canada}


\begin{abstract}
The confrontation of the Cosmic Microwave Background (CMB) theoretical
angular
power spectrum with available data often
requires the calculation of large numbers of power spectra.  The standard
practice is to use a fast code to compute the CMB power spectra over some
large parameter space, in order to estimate likelihoods and constrain
these parameters.
But as the dimensionality of the space under study increases, then even
with relatively fast anisotropy codes, the computation can become
prohibitive.
This paper describes the employment of a `morphing' strategy to
interpolate new power spectra based on previously calculated ones.
We simply present the basic idea here, and illustrate with a few
examples; optimization of interpolation schemes will depend on the
specific application.  In addition to
facilitating the exploration of large parameter spaces,
this morphing technique may be helpful for Fisher
matrix calculations involving derivatives.
\end{abstract}

\end{frontmatter}


\section{Introduction}

Detailed measurement of the anisotropies on the Cosmic Microwave
Background
(CMB) sky promises to reveal a wealth of information about the Universe
in which we live.  Spurred on by the rapid advances in CMB
experimentation and the promise of the \MAP\
(\url{http://map.gsfc.nasa.gov}) and \Planck\ 
({\url{http://astro.estec.esa.nl/SA-general/Projects/Planck/}) satellites,
there has also been great activity in recent years focussed on CMB theory
and data analysis \citeaffixed{Bond97}{e.g.}.  Because of the enormous
size of
future data sets, anything which might speed up the task of extracting the
full cosmological information could be extremely useful.

The power spectrum of CMB anisotropies is usually expressed in terms of
the multipole moments $C_\ell$, which are the expectation values of the
squares in a spherical harmonic expansion of temperatures on the sky.
Here $\ell$ is an inverse angle, and then $C_\ell$ vs $\ell$ is just a
plot of the power spectrum of fluctuations, analogous to $P(k)$ vs $k$
for a power spectrum derived from Fourier modes in flat space.
It is conventional to plot $\ell(\ell+1)C_\ell$ (which is the power per
decade in $\ell$, and which we will refer to as
$\mathcal{C}_\ell$) vs $\ell$, analogous to $k^2P(k)$ vs $k$ for a flat sky.
For a given set of cosmological parameters, the theoretical prediction for
the $\mathcal{C}_\ell$ curve can be calculated quite precisely
\citeaffixed{HSSW}{see e.g.}.  For all
popular cosmologies one finds a series of bumps and wiggles.  It is this
rich structure which promises to reveal the values of the cosmological
parameters, and which also allows us to utilize the interpolation method
described below.

When data from the \COBE\ satellite first became available
\cite{Smoot92}
there were a large
number of attempts to fit to cosmological models.  But the restricted
range of angles probed by the \COBE\ beam meant that the data were really
only sensitive to an amplitude and slope, with some mild constraint on
the curvature \cite{BunWhi}.
Therefore the suite of models which needed to be considered was relatively
modest.  As more data from smaller scales became available, probing the
acoustic peak region, there were early attempts to constrain a broader
range
of models \citeaffixed{ScoWhi,BondDahlem}{e.g.}.  As even more data
poured
in it soon became clear that it was necessary to search a parameter space
with a significant number of dimensions
\citeaffixed{Hanetal,Lineweaver,BonJafKno,Teg99,Efsetal,DodKno}{e.g.}.
Some of these studies have attempted to fit
likelihoods for models calculated in as many as 6 separate parameter
dimensions (or even more using relations between parameters, or other
tricks to reduce the dimensionality).

Even larger, and higher quality data sets are expected in future.
Long-duration balloon flights such as BOOMERANG are already producing
data, and
at least three dedicated interferometers are under development.  The \MAP\
satellite is due for launch in 2000, and the \Planck\ mission is due for
launch in 2007.  Such data sets will require significantly more thorough
exploration of the available parameter space in order to extract the most
accurate information about our Universe.  In practice this will involve
the
calculation of truly vast databases of theoretical models.  As a result, 
methods of rapidly obtaining the power spectra are crucial.  Codes have
been developed \citeaffixed{SelZar96}{the most widely used being
{\tt cmbfast},} which efficiently and accurately
calculate the anisotropy power spectra.  Nevertheless these codes still
take
a significant amount of time when one considers that there may be a
roughly
10 dimensional parameter space to explore.  Therefore it is worth
considering whether there are any short-cuts which can be taken, with
minimal loss of accuracy, for exploring the full parameter space.
Some form of accurate interpolation would be
particularly useful, since it is orders of magnitude faster to interpolate
existing curves than to generate new ones.

The CMB power spectra curves are quite smooth,
and vary smoothly with individual parameters,
suggesting that it may be unnecessary to perform explicit calculations for
every value of a particular parameter -- for a reasonable level of
accuracy, some type of interpolation is probably sufficient.

The obvious interpolation scheme is simple linear interpolation in the
vertical direction.  However, given that the curves contain a handful of
special features which deform quite smoothly,
we can easily imagine interpolation schemes which
more explicitly involve continuous changes of one curve into another.
Inspired by an idea from image manipulation, sometimes called
`morphing',\footnote{Definition: `The animated transformation of one
image into another by gradually distorting the first image so as to move
certain chosen points to the position of corresponding points in the
second
image' (from the Free On-Line Dictionary of Computing,
\url{http://foldoc.doc.ic.ac.uk}).}
we explore the use of such an approach for interpolation of
CMB power spectra.  We are aware of at least one other application of
similar ideas in astrophysics, and that is in the generation of stellar
isochrones from a relatively small number of stellar mass models,
as first applied by \citeasnoun{Prather}, and described in detail by
\citeasnoun{BerVan}.

\section{Description}

The underlying method behind any type of morphing scheme involves
selecting certain special features of the images being morphed, defining
how these features will be mapped into each other, and then using these
mappings to create smooth transformations between the other parts of
the objects \citeaffixed{Wolberg,MagTha,Gometal}{see e.g.}.
Morphing is just the computerized version of a process which is
historically well-known in cinematic animation, where a set of
`key frames' is generated by skilled artists, and then the
process of `in-betweening', or filling in the steps between these frames
is performed by less skilled artists.  A common example used to display
the effectiveness of morphing in computer graphics is that of two
different human faces being morphed into each other, passing continuously
through a composite face.  In this case the special
features might be the eyes, nose, mouth, and cheekbones, and these
features might be mapped into each other using linear interpolation along
a straight line between corresponding points of the picture.  A smooth
mapping field for the rest of the picture could then be generated by
interpolating between these mapping vectors.

The procedure is very much the same when morphing is used to smoothly
interpolate between two similar curves.  The difference in this case is
that
usually each curve is associated with a vector of parameters, and we wish
to accurately interpolate the intermediate curve associated with some
intermediate parameter vector (corresponding to `tweening' in animation).
Analogous to the eyes or nose of a face, 
physically or mathematically significant
points on the curve are chosen to be the special features which are mapped
into each other -- these are known as the morphing `control points'. The
locations of these control points on the intermediate
curve -- which we call `target points' -- is interpolated using the
control
points from the original curves and the associated parameter vectors.
Once
the locations of these target
points has been determined, a vector is constructed between each control
point
and the corresponding target point.  By interpolating between these
vectors a
mapping field is created that will morph the curves to the shape
approximating the curve at any intermediate location in the parameter
space.

Having understood the general principle, the question is:
can the CMB angular power spectrum be interpolated using some type
of morphing algorithm?  The fact that it changes
smoothly and continuously 
as cosmological parameters are varied suggests that the answer is `yes'.
We give some examples to illustrate this below.
 
At this point it is worth stressing why this morphing approach is superior
to linear interpolation.  The basic reason is that morphing involves
interpolating the curves 
{\it in a coordinate system which is matched to the physical process\/}
which is distorting the curves.
A good analogy is to consider how to interpolate between two circles of
different radii.  To make the problem single-valued, consider only
the semi-circles with positive $y$-coordinate.  Now imagine trying to
interpolate to another semi-circle with an intermediate radius.
It is obvious
that linear interpolation in the vertical coordinate is hopeless, and that
switching to polar, and interpolating radially,
is the sensible approach.  The point about
morphing, using special points on the curves, is that one is effectively
using the coordinate system in which the curves are naturally changing.
So once we interpolate the curves in this `morphological coordinate
system', then we can perform interpolation in the usual way, except that
we have now used the shape information of the curves to define the
direction in which to interpolate.

\section{Details}
\label{sec:details}
We now describe each step in the morphing procedure.
We will assume that a set of curves covering the range of parameters of
interest has already been calculated, and we are
attempting to interpolate a `target curve' at some intermediate parameter
value (see Fig.~\ref{fig:control}).

Note that for many of these
steps several different choices could be made.  In no case do we claim
that what we describe below represents the optimal choice.
Rather, we hope that our outline of the algorithm is detailed
and clear enough to stimulate other studies.

\begin{figure}
\begin{center}
\resizebox{\hsize}{!}{\includegraphics{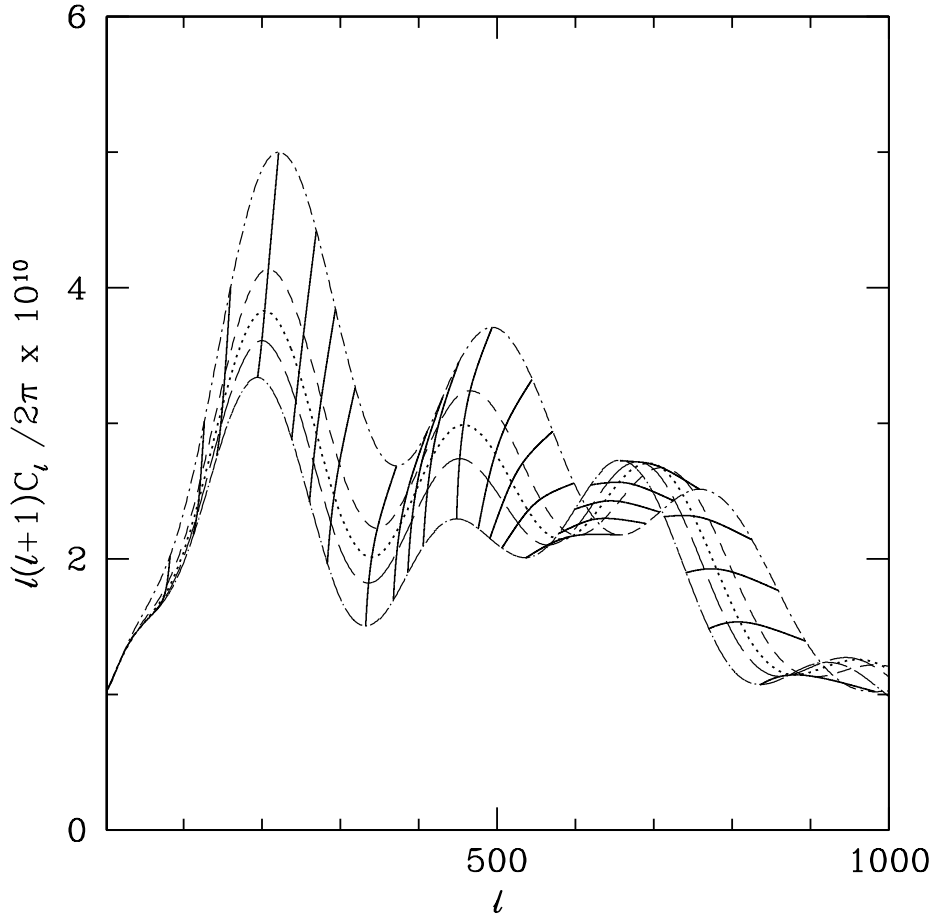}}
\end{center}
\caption{
This figure shows how 4 point cubic splines can be used interpolate the
target points for CMB anisotropy power spectra.
Dashed and dot-dash lines represent 4 different
cosmological models (actually different values of the Hubble constant).
For definiteness we chose standard Cold Dark Matter type models, and
concentrated on multipoles $\ell\,{=}\,2$--1000.
The solid lines indicate spline fits to the positions of primary and
secondary control points.  The dotted line shows the interpolated model.
}
\label{fig:control}
\end{figure}

\subsection{Choose control points}

The first step in any morphing process is to choose which points will be
used as the morphing `control points'.
Morphing control points are special points on the curves defining the path
through which morphing occurs
as a function of the interpolation parameter.

It is helpful conceptually to separate the control points into two
classes, primary control points and secondary control points.  Primary
control points are easily identifiable features on the curves, and
track gross changes in the shape of the curve, while secondary
control points are selected between the primary points and
track the higher order effects.  The criteria
for good primary control points are that they are morphologically
significant, and easy to determine automatically.

The obvious choice for the primary 
control points for the CMB power spectra are the maxima and minima, 
which are also significant in that they are related to the underlying
physical processes.  Because the CMB power spectrum is relatively simple
and smooth, a good choice for the secondary control points might be the
inflection points of the curve, i.e.~the extrema of the derivatives.
However, once the primary control points have been selected, there are
many 
options for secondary control points, because they are only tracking small
deviations in the shape of the curve.  Another class of points are those
which lie some definite fraction of the way vertically (or horizontally)
between a maximum and a minimum.  One advantage of these points is that 
they are unambiguously unique, owing to their construction between a
maximum and a minimum.  We have found that these points may be a better
choice for secondary control points, since inflection points or the
extrema of higher order derivatives are numerically
difficult to find with sufficient accuracy.  In the examples which follow,
we have used as secondary points those points
$\fraction1/4, \fraction1/2$ and $\fraction3/4$
of the vertical distance between the primary points.

There is also a question of what to use as control points at the lowest
and highest multipoles considered.  In our examples we simply chose
$\ell\,{=}\,2$ and $\ell\,{=}\,1500$ as control points, which is certainly
not optimal -- it would be better to choose something more physical than
just the end-points.  In practice, however, it is less important to be
fully accurate at the lowest $\ell$s, where cosmic variance is large, and
at the very highest $\ell$s, where the primordial spectrum falls off.

\subsection{Find control points}

Once the control points have been selected they must be extracted from
each curve.  For the purposes of morphing it is useful to consider the
CMB power spectrum as a continuous function of $\ell$, for which we only
have samples at integer $\ell$.  Using standard routines it is
straightforward to determine the maxima and minima of this function.
Finding secondary control points based on these maxima and minima is
equally straightforward.  If we wanted instead to determine the
inflection points of this function, a derivative would need to be
calculated at this stage.  Precautions must be taken 
to ensure the original functions are free of high frequency noise prior to
this step, as unwanted inflection points induced by noise introduce an 
ambiguity that can play havoc when mapping the control points.  
In practice we low-pass filtered when we examined
numerical differences -- but even so -- we found that it was difficult to
accurately pin down turning points in the derivatives of
the $\mathcal{C}_\ell$s.

\subsection{Calculate the target points}

The next part of the procedure involves finding the points on the
intermediate curve which correspond to the control points -- we call
these the `target points'. 

In fact the key step that governs how accurately an intermediate set of
$\mathcal{C}_\ell$s can be determined is how accurately the target
points can be interpolated from the control points.  The simplest method
would be to linearly 
interpolate between sets of control points on two existing curves.
The next simplest method
would be to use the curves from several points in parameter space instead
of just two, and to use spline interpolation
to obtain the target points from the sets of control points.
Although the lowest-order cubic spline requires more (actually 4)
sets of $\mathcal{C}_\ell$s compared with linear interpolation,
the accuracy of the method is significantly improved (as described
later).  We therefore adopted 
4 point cubic spline interpolation to find the target points
in the morphing studies described here.

\subsection{Map control points}

{From} the point of view of writing a code to automate the morphing
process, the thorniest step is mapping one set of control points into
another set.  The morphology of each $\mathcal{C}_\ell$ curve can be 
classified based on the 
set of primary control points extracted from it. Even though the CMB power
spectrum smoothly and continuously changes shape, the morphology can and
does change {\it discontinuously\/} at some parameter values.
There are two major ways that the morphology can change from one set of 
parameters to another.  The first is an artifact of simulating the power
spectrum over a finite range of $\ell$s, and occurs when a minimum or
maximum enters or leaves the edge of the simulated range.  The second
reflects an actual change in the morphology of the curve, and occurs when
a maximum and minimum simultaneously disappear or appear on the curve at
some intermediate location.  The tricky part of mapping control points is
detecting which combination of morphological changes has occurred, and
then mapping the control points accordingly.  Each $\mathcal{C}_\ell$
curve could
also be classified at a more detailed level based on the secondary control
points if they 
have some non-uniform structure, however this would further complicate the
mapping algorithm.

Successfully mapped control points are used to
generate the target points for the warping stage.
Although a change in morphology will leave some control points unmapped
-- and thus unwarped  -- the final step in the process will interpolate 
these regions accurately.

Exactly how versatile this mapping process needs to be obviously depends
strongly on the parameter range being explored.  In practice one could 
calculate in advance the places in parameter space where maxima and minima
converge, and be prepared for this.  Ultimately this may not even work,
and for some parameter ranges, it may
be necessary for the interpolation algorithm simply to instruct the user
to provide further $\mathcal{C}_\ell$ curves in order for accurate
interpolation to be achievable.
            
\subsection{Warp curves}

After the control points have been mapped, and the target points
generated, each of the curves nearest to the desired location in
parameter space are `warped' into the expected shape.
A nice way to envision the warping is to separate it into two steps.
First, the curve is re-parameterized on the horizontal axis to a 
coordinate system where the control points are aligned vertically at
the target $\ell$ value.  Secondly,
the curves are then cubic spline interpolated in the vertical direction, 
constraining the warped curve to pass through the target points.

\begin{figure}
\resizebox{\hsize}{!}{\includegraphics{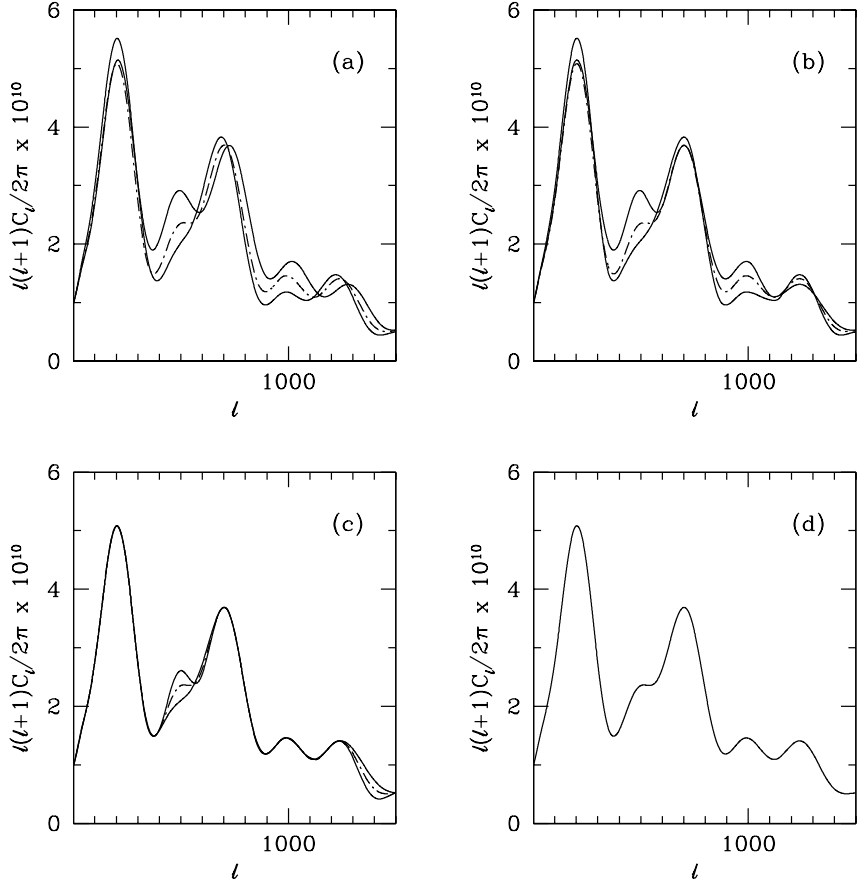}}
\caption{
Demonstration of the method with three sets of computed $C_\ell$s,
one at a parameter value intermediate to the other two (dot-dash curve):
(a) starting point; (b) after horizontal warping;
(c) after vertical warping as well; (d) after weighted
averaging of the two warped curves.  In the final panel the interpolated
(solid) curve covers the target (dot-dashed) curve almost exactly.
}
\label{fig:warpdemo}
\end{figure}

In practice it may be easier to consider each $\mathcal{C}_\ell$ curve to be a
vector function $\vec{f}$.  From this point of view, at the $k$th control
point a warp-vector is created that will move the control point into the
target point, i.e.

\begin{equation}
\vec{f}_k \equiv [\Delta\ell_k,\Delta (\mathcal{C}_\ell)_k],
\end{equation}
where
\begin{equation}
\Delta \ell_k \equiv \ell^{\rm target}_k - 
\ell_k, 
\end{equation}
and
\begin{equation}
\Delta (\mathcal{C}_\ell)_k \equiv
 {(\mathcal{C}_\ell)}^{\rm target}_k - (\mathcal{C}_\ell)_k.
\end{equation}

A spline is used to interpolate between these vectors, generating a smooth 
warp-vector function $\Delta \vec{f}$.  By applying this type of function
to both of the nearby $\mathcal{C}_\ell$ curves, two warped curves are
generated, $(\mathcal{C}^1_\ell)^{\prime}$ and
$(\mathcal{C}^2_\ell)^{\prime}$,  that approximate the 
curve at the desired location in parameter space:

\begin{equation}
\vec{f} \longrightarrow \vec{f}^{\prime} = \vec{f} + \Delta \vec{f},
\end{equation}
\begin{equation}
{[\ell,(\mathcal{C}_\ell)] \longrightarrow
 [\ell^{\prime},(\mathcal{C}_\ell)^{\prime}] = [\ell
+\Delta\ell, (\mathcal{C}_\ell) + \Delta (\mathcal{C}_\ell)]}.
\end{equation}

\subsection{Obtain the final curve}

The penultimate step in the process is to re-grid the warped
$\mathcal{C}_\ell$
curves onto integer values of $\ell$ again, and calculate
the weighted average of the two warped curves at each $\ell$, i.e.

\begin{equation}
(\mathcal{C}_\ell)^{\rm target} =
 x (\mathcal{C}^1_\ell)^{\prime} + (1-x) (\mathcal{C}^2_\ell)^{\prime},
\end{equation}
where $0\,{<}\,x\,{<}\,1$.  For example if we are interpolating the curve
at a parameter
value exactly half-way between the value at the two
curves, then $x\,{=}\,\fraction1/2$
and we have the simple average of the warped curves.
This final step ensures that the transition between curves will be
continuous,
and will complete the morphing process in regions near morphological 
discontinuities.  A step-by-step pictorial summary of the morphing 
process is shown in Fig.~\ref{fig:warpdemo}.

\section{Accuracy}

\begin{figure}
\resizebox{\hsize}{!}{\includegraphics{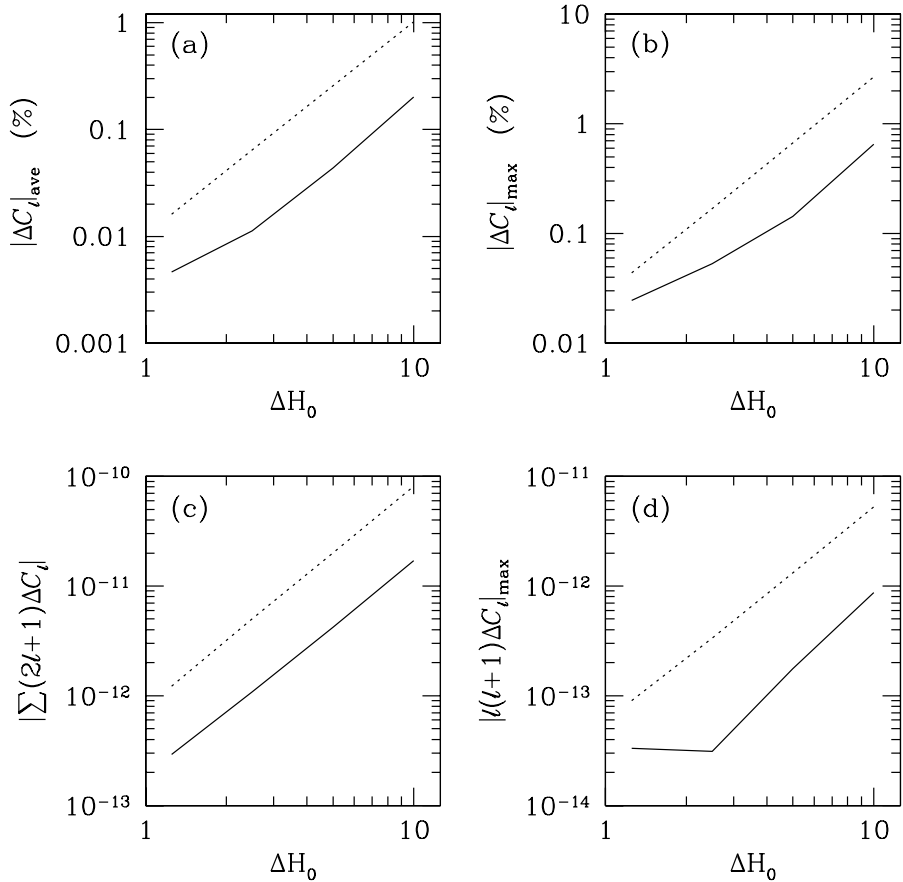}}
\caption{
Quantification of error for one example of morphing interpolation, in the
Hubble constant, shown as a function of the grid spacing in $H_0$:
(a) mean percentage error; (b) maximum percentage error;
(c) deviation in total power; (d) maximum absolute deviation.
The dotted line represents linear interpolation, while the
solid line is for morphing.
}
\label{fig:hubble}
\end{figure}

There are various places where the details of our approach are probably
not
optimal, and where further study could be done.  For all subsections of
Section~\ref{sec:details}
there are possible refinements which might improve the accuracy.
Examples include: choice of primary control points and number of secondary
control points; numerical method of finding these points; interpolation
procedure for obtaining the target points, including the number of curves
used and the method of interpolating between them; the detailed procedure
for interpolating between the warp-vectors; and the method of weighting
the two warped curves to obtain the target curve.
The best method may depend on details of the problem being addressed,
e.g.~the parameter
range, value of $\ell_{\rm max}$ etc.  Indeed, even the statistic for
describing how well the interpolation has performed might depend on the
specific application.  For example, should the error be weighted by the
inverse of the cosmic variance?

There are several statistics that we felt might be relevant when
considering
the accuracy of this method, including maximum error, location of 
maximum error, average error, and error in the total power
($\sum(2\ell+1)\Delta C_\ell$).  These statistics (actually their
absolute values) are plotted for several
cases in Fig.~\ref{fig:hubble}.  Another possibility is to consider the
deviation in the cummulants of the curves, i.e.
\begin{equation}
\max\left\{ \sum_{\ell=2}^{k} \left(C_\ell^1-C_\ell^2\right) \right\},
\end{equation}
which gives similar results.

Fig.~\ref{fig:hubble}
is for the specific example of interpolation in $H_0$, around the
parameter values of the standard Cold Dark Matter model.  We see that
morphing
reduces the maximum percentage error, average percentage error, total
power
error, and the maximum absolute error
all by a factor ${\sim}\,5$ over vertical linear interpolation.
Note that as the grid spacing is reduced the error functions start to
converge, due to the noise limitations of specific runs of
the {\tt cmbfast} code which we used.

\begin{figure}
\resizebox{\hsize}{!}{\includegraphics{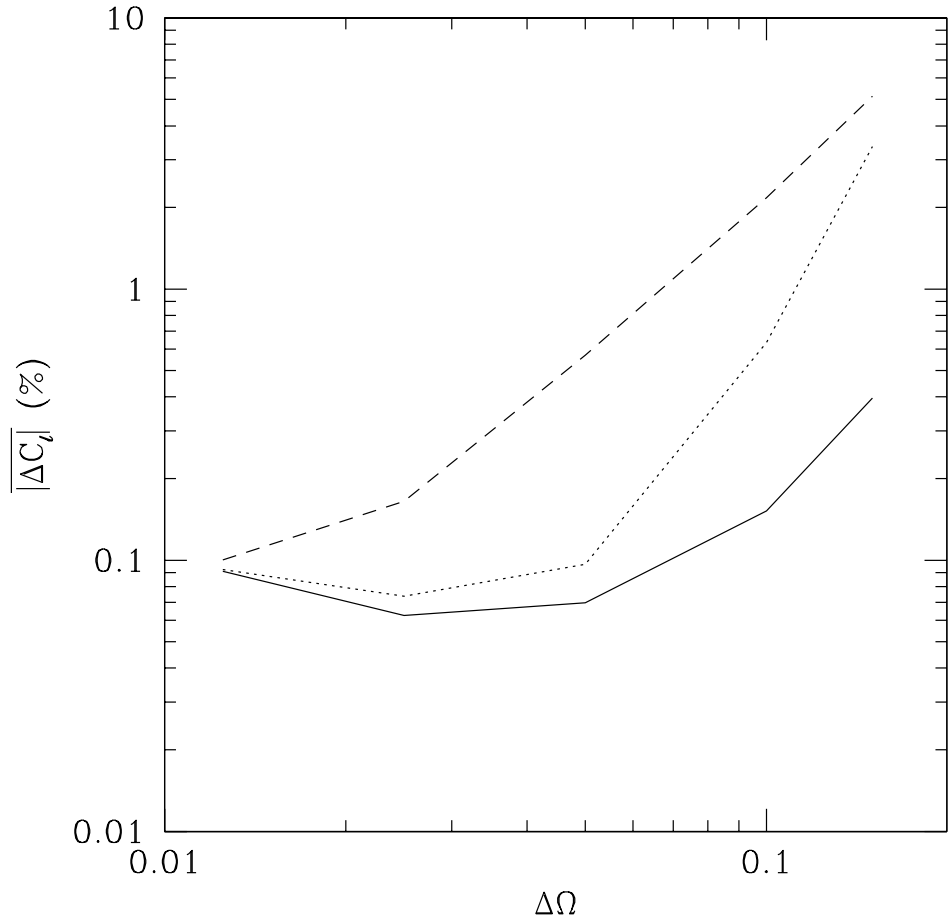}}
\caption{
Mean percentage error as a function of the grid spacing in $\Omega_0$.
The dashed line shows linear interpolation, while the dotted line shows
cubic spline interpolation in the vertical direction, and
the solid line is for morphing.
}
\label{fig:omega}
\end{figure}

Now let us look at morphing in another parameter.
As $\Omega_0$ (the density parameter, assuming here that there
is no cosmological constant) is decreased, then to a first approximation
the location of the peaks of the power spectra are simply shifted to
higher values of ${\ell}$.  Because of this, we expect that morphing
on $\Omega_0$ should be far superior to vertical interpolation on
$\Omega_0$.  Indeed, in Fig.~\ref{fig:omega} we see
that, for large grid spacing, morphing reduces the average absolute
percentage error by a factor of ${\sim}\,13$ compared to vertical linear
interpolation.  And even if we used cubic spline interpolation in the
vertical direction, we find that morphing is still better by a factor of
${\sim}\,8$.
Again, as the grid
spacing in $\Omega_0$ is reduced all three curves rapidly converge to
the level of accuracy with which the model $\mathcal{C}_\ell$s were
generated.  However, if the original $\mathcal{C}_\ell$s are calculated
more precisely, then morphing
continues to give accurate results down to smaller parameter spacing.

In practice morphing will be used on a grid of several parameters at once.
Multi-dimensional interpolation is complicated by the fact that
the final answer usually depends on the order of the parameters one 
interpolates with.
Therefore,
another useful comparison is to perform morphing on two different
parameters in the two possible orders, and examine how closely the end
results match.  This tests the linearity of the method,
and gives another estimate of its accuracy.  As an example we examine
morphing for the baryon fraction $\Omega_{\rm b}$ and the dimensionless
Hubble parameter $h$.  For typical
grid spacings, Fig.~\ref{fig:twodim} indicates that the maximum deviation
between the two possible 
permutations is on the order of 0.1\%, with the average deviation
an order of magnitude less.

\begin{figure}
\resizebox{\hsize}{!}{\includegraphics{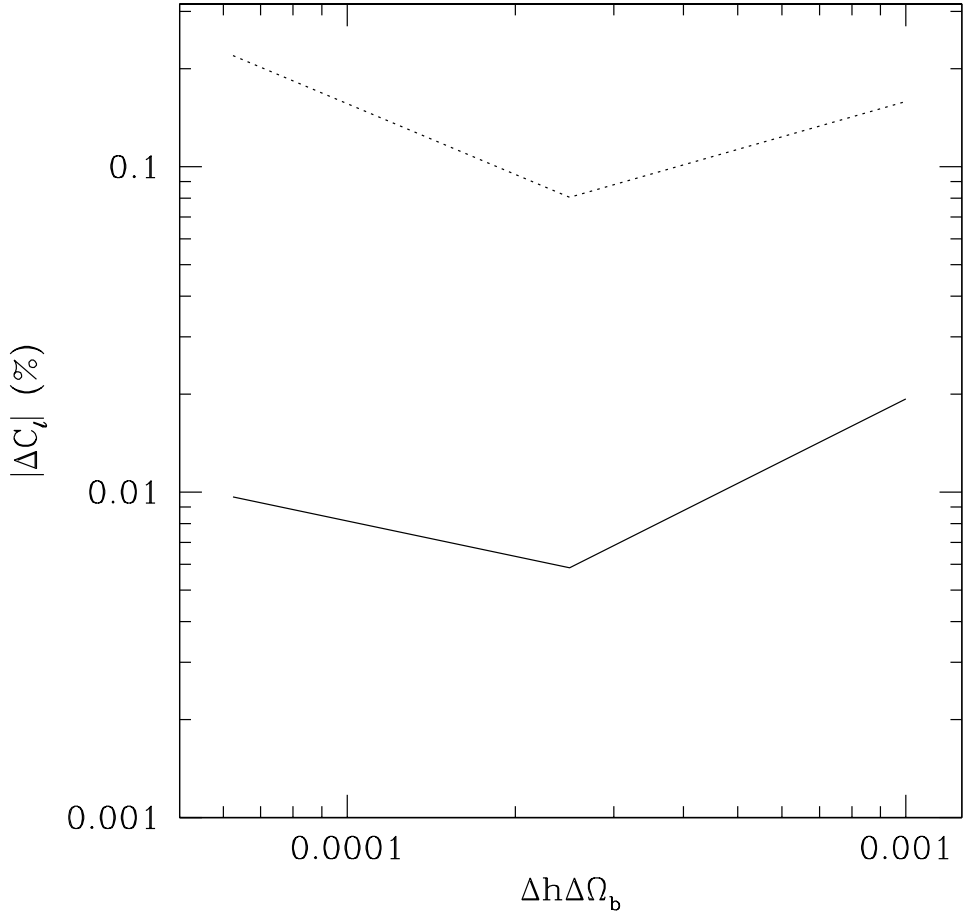}}
\caption{
The difference between the two possible permutations of a two dimensional
morph on the parameters $h$ and $\Omega_{\rm b}$,
plotted as a function of the unit grid area.
The solid line indicates the average deviation, while
the dotted line shows the maximum deviation. 
}
\label{fig:twodim}
\end{figure}

Interpolation between the models is inherently several orders of magnitude
faster than direct calculation.  While this algorithm does have several
extra steps compared with simple vertical interpolation, the actual runtime
for morphing is comparable.  It is certainly essentially
instantaneous compared to direct calculation of the $\mathcal{C}_\ell$s. 

\section{Discussion}

It is clear that vertical interpolation is optimized when the major
changes in the shape of the curve occur vertically.  The major advantage
of morphing over simple vertical linear (or cubic spline)
interpolation is the ability to accurately track horizontal distortions of
a curve as well.
Morphing is particularly useful in regions of the curve where the
slope 
is steep, as these small horizontal shifts in the curve will result in
large
vertical shifts at a fixed horizontal coordinate.
Morphing can track these changes more accurately because when
morphing, we effectively change to a coordinate system which tracks the
major changes in the shape of the curve -- the `morphological
coordinate system'.

Let us introduce a parameter, $\lambda$, which tracks whether the curves
are
being interpolated vertically in the initial coordinate system, or along
vectors suggested by the control points (vertically in the morphological
coordinate system) -- $\lambda$ can be any smooth function, such as an
angle, describing the interpolation direction.  For simplicity we will
define $\lambda\,{=}\,0$
to be simple vertical interpolation, $\lambda\,{=}\,1$ is the morphing
direction, and $\lambda\,{>}\,1$ is interpolation along vectors which are
even more horizontal than that.  Explicitly we can take $\lambda$ to be
the fraction of horizontal transformation that we use, i.e.~take
\begin{equation}
        \ell \to \ell^\prime = \ell+\lambda\cdot\Delta\ell,
\end{equation}
where $\Delta\ell\,{=}\,\ell^{\rm target}-\ell$.
A good indicator of the relative accuracy that can be achieved by vertical
interpolation between two well behaved curves is their mean normalized
absolute difference:
\begin{equation}
\left\langle\Delta_N\right\rangle
 \equiv \left\langle{\frac{|f_1-f_2|}{\min[f_1,f_2]}}\right\rangle.
\end{equation}
A lower value of this statistic indicates that the average normalized
vertical distance between the two curves is less, and consequently the
magnitude of the errors induced by vertical interpolation is nearly always
reduced.

Fig.~\ref{fig:mystery} shows an example of a calculation of this
statistic for a particular CMB power spectrum interpolation.  What we see
is that as we continuously transform from
the original coordinates to the morphological coordinates, $\Delta_N$
goes through a minimum near $\lambda\,{=}\,1$, the value of $\lambda$ in
the
morphological coordinate system.  This graph indicates that by changing to
the morphological coordinate system we expect the errors introduced by
vertical interpolation to be reduced.
Because the errors introduced by transforming to the morphological
coordinate system are relatively small, the overall effect of morphing is
to reduce interpolation errors.

\begin{figure}
\resizebox{\hsize}{!}{\includegraphics{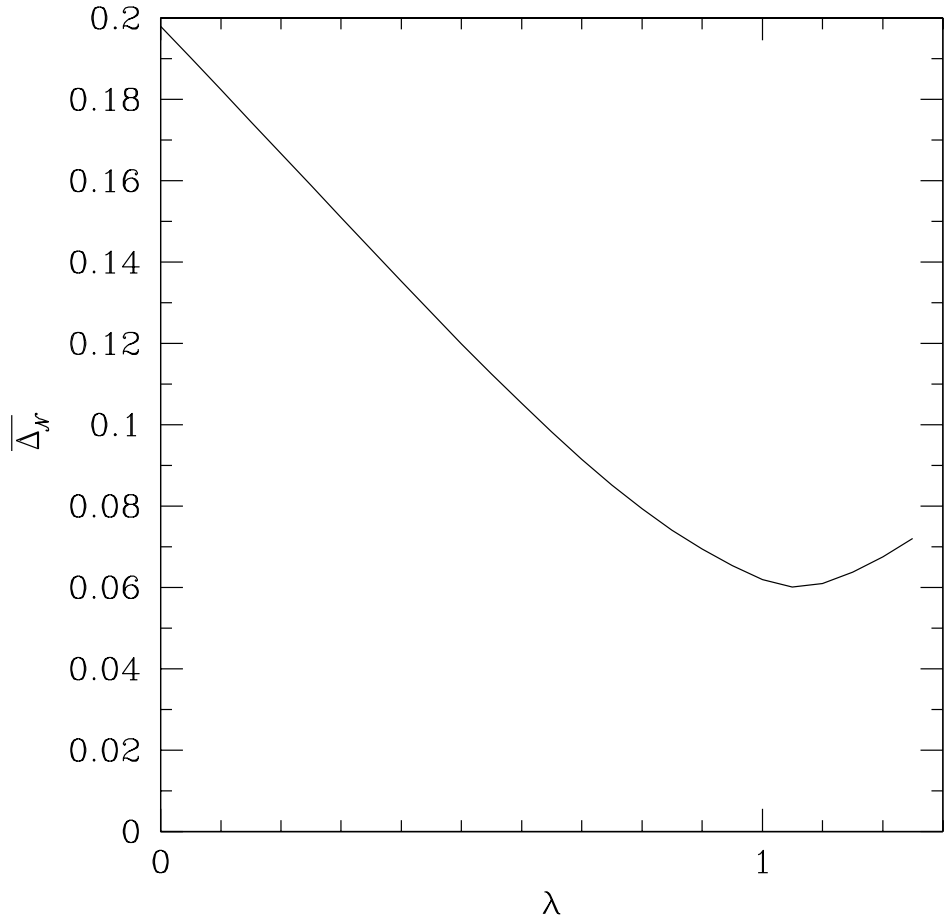}}
\caption{
The mean normalized difference between the nearest interpolating
curves is shown as a function $\lambda$, a parameter which tracks the
transformation to the morphological coordinate system.  $\lambda\,{=}\,0$
represents the original coordinate system (i.e.~vertical interpolation),
$\lambda\,{=}\,1$ represents a full transformation to morphological
coordinates (i.e.~interpolating along the vector separating the control
points), and $\lambda,{>}\,1$ represents an over-transformation
(i.e.~further from the vertical than even the morphing direction). 
}
\label{fig:mystery}
\end{figure}

The process of morphing keeps track of the most important information about
a function from a numerical standpoint.  If the locations and values of the
maxima and minima of a function and several derivatives are known, we  can
easily reconstruct a very good approximation to that function.  Since often
the important physical information we can extract from a curve is encoded
in these prominent features of the curve, it makes sense to devote most
effort to interpolation of these features.  In fact, it may be fruitful to
look into the possibility of replacing the thousands of $\mathcal{C}_\ell$s
with a smaller vector of numbers that keeps track of the prominent features
of the curve.  If we stop the morphing algorithm after the control points
have been determined, we effectively accomplish this `compression' of the
critical information in the power spectrum (the advantage of our algorithm
is that by smoothly warping the curves we use information from the regions
{\it between\/} the control points to reduce inaccuracies).  One can easily
imagine trying to reconstruct the power spectrum using {\it only\/} the
control points -- at the cost of additional inaccuracy.  Alternatively, one
could attempt to directly fit the positions of the control points as a
function of the cosmological parameters.  We leave such investigations to
future studies.

The most troublesome instabilities in the morphing algorithm arise because
of the discrete nature of the power spectrum.  Near to morphological
discontinuities the algorithm may try to explode a small region around 
the discontinuity, of say $\Delta\ell\simeq10$, to cover a much larger
span in $\ell$.  This results in an excessively flat function over this
region that dramatically increases the error.   The solution to this
problem is to create a `buffer zone' around regions of the
$\mathcal{C}_\ell$ curve near morphological
discontinuities in parameter space, to prevent points in these regions
from
being used as control points. 

Other improvements are probably possible.
We have not explored whether additional physical information, such as
known approximate dependencies on parameters in certain regions of the
parameter space, would significantly improve the
interpolations.  For example, we can certainly imagine that using a power
law in $\Omega_{\rm K}$ (the curvature) might be a better way of
interpolating between the control points than the spline fit that we have
used.  The positions and heights of the peaks can be understood
analytically \citeaffixed{HuSugSil}{e.g.} through the physics of acoustic
modes driven by gravity.  Hence it might be possible to replace the spline
interpolation step of the morphing algorithm with a function which is more
physically motivated.  Another refinement might be to include a control
point
related to the curvature scale at low $\ell$ in open or closed models
(or the angular scale corresponding to the epoch when potentials
are decaying in models with a cosmological constant).  Further
elaborations of morphing will depend on the parameter ranges being
considered.

\section{Applications}

The most obvious application for morphing is in the construction of
large likelihood grids for the analysis of future CMB data sets.
If every $M$th grid point is calculated
by direct means, morphing reduces the calculation time by a factor of
nearly $M^N$, where $N$ is the dimensionality of the likelihood
space being explored. 
While this fact is also true for simpler forms of interpolation, the
increased accuracy of morphing allows for larger values of $M$ while
maintaining the same level of approximation.
Techniques like this will be crucial for the searches in ${\sim}\,10$
parameters that will soon be necessary.

Another possible application of the increased accuracy of morphing is when 
calculating derivatives of the $\mathcal{C}_\ell$s with respect to the
model parameters.  These derivatives are useful for Fisher matrix studies
\citeaffixed{EisHuTeg}{e.g.}, for instance.  Morphed interpolation allows
for much smoother and slightly more precise estimates of the derivatives.

\section{Conclusions}

Morph-interpolation gives a way to confront the problem of exploring
large likelihood spaces to a satisfactory degree of accuracy.  The
superiority of morphing to simple vertical interpolation is apparent.
This superiority comes at the cost of a modest increase in coding
complexity,
but little increase in computation time. 

Morphing has recently become an integral part of the
modern computer graphics artist's repertoire.  We hope that this
 paper will stimulate
interest in morphing as an approach to interpolation,
and that it might also encourage
people to look for inspiration outside of astrophysics when tackling
new problems such as those associated with the confrontation of CMB data
with
theoretical models.

\ack{}
This research was supported by the Natural Sciences and Engineering
Research Council of Canada.  The code {\tt cmbfast}, developed by
Uro{\u s} Seljak and Matias Zaldarriaga, was used to generate the
$C_\ell$s.


\end{document}